\documentclass[reprint,aps,prb,floatfix]{revtex4-2}
\usepackage{bm}
\usepackage{amssymb}
\usepackage{amsmath}
\usepackage{graphicx}
\usepackage{rotating}
\usepackage{epsfig}
\usepackage{psfrag}
\usepackage{amsmath}
\usepackage{subfigure}
\usepackage{braket}
\usepackage{tikz}
\usepackage{stmaryrd}
\usepackage{wasysym}

\newcommand{\bea}{\begin{eqnarray}}
\newcommand{\eea}{\end{eqnarray}}
\newcommand{\bpm}{\begin{pmatrix}}
\newcommand{\epm}{\end{pmatrix}}

\usepackage[unicode=true,colorlinks=true,pdfpagemode=UseOutlines,pdfstartview=Fith]{hyperref}
\hypersetup{linkcolor=blue,citecolor=blue,urlcolor=blue}

\newcommand{\hz}[1]{\textcolor{blue}{#1}}

\begin{document}
\title{Electric field driven spin textures in heavy fermion van der Waals magnets}

\author{Aayush Vijayvargia$^1$, Hao Zhang$^2$, Kipton Barros$^2$, Shi-Zeng Lin$^{2,3}$, Onur Erten$^1$}
\affiliation{$^1$Department of Physics, Arizona State University, Tempe, AZ 85287, USA\\ $^2$Theoretical Division and CNLS, Los Alamos National Laboratory, Los Alamos, New Mexico 87545, USA \\ $^3$Center for Integrated Nanotechnologies (CINT), Los Alamos National Laboratory, Los Alamos, New Mexico 87545, USA}

\begin{abstract}
The recently discovered van der Waals material CeSiI exhibits both heavy fermion behavior and spiral order with strong magnetic anisotropy which makes it a potential host for topological spin textures such as skyrmions through electrical gating. A monolayer of CeSiI consists of two layers of Ce atoms on triangular lattices that sandwich a silicene layer. Motivated by the experiments, we explore magnetic phase diagram in van der Waals heavy fermion materials as a function of anisotropy and applied magnetic field using an effective spin model. We demonstrate that application of an external electric field can tune the Kondo coupling on each Ce layer differently, in turn allowing for controlling the intra- and interlayer magnetic couplings. Our analysis indicates that this fine-tuning leads to the coexistence of different magnetic orders in a single monolayer. In particular, we show that a novel vortex phase can be stabilized only in the presence of an external electric field. Our results highlight the unique advantages and the tunability of van der Waals heavy fermion materials for manipulation of chiral magnetic phases.
\end{abstract}
\maketitle
\section{Introduction}
Monolayers of magnetic van der Waals (vdW) materials exhibit a variety of orders such as ferromagnets, antiferromagnets, spirals and skyrmions\cite{Blei_APR2021}. Heterostructures and moir\'e superlattices of vdW magnets can further lead to novel spin textures that may not be possible to realize in monolayers\cite{Hejazi_PNAS2020, Hejazi_PRB2021, Akram_PRB2021, Akram_NanoLett2021, Akram_NanoLett2024, Xu_NatNano2022, Song_Science2021, Xie_NatPhys2023}. While a majority of vdW magnets are Mott insulators, the recently discovered CeSiI exhibits both magnetic order and heavy fermion behavior\cite{Posey_Nature2024}. Unlike traditional intermetallic heavy-fermion compounds, CeSiI consists of 2D metallic sheets weakly held together by vdW interactions. This allows the sheets to be cleaved and the heavy fermion behaviour has been established down to 2D limit\cite{Posey_Nature2024}. The 2D limit is desirable as it allows for controlling its properties via external knobs such as gating, electric field, strain\cite{Song_arXiv2024}. Building heterostructures of these vdW materials aided with these new tuning knobs allows exploration of even richer physics that may not be possible in the monolayer or the bulk.

Skyrmions are topological defects of 2D magnets with complex non-coplanar spin textures. First discovered in B20 compounds\cite{Binz_Science2009},  skyrmion crystals (SkX) arise as superposition of three magnetic spiral states. These spirals may be realized as a results of the competition between ferromagnetic exchange and Dzyaloshinskii-Moriya interactions due to the inversion symmetry breaking. Alternatively they can arise in frustrated systems with competing interactions such as a triangular lattice with ferromagnetic nearest neighbour (NN) and antiferromagnetic next-nearest neighbour (NNN) interactions\cite{Okubo_PRL2012}. The spiral states can transform into a SkX under applied magnetic field and an moderate easy axis anisotropy\cite{Lin_PRB2015, Banerjee_RPX2014, leonov_natcomms2015, lin2016ginzburg, hayami2016bubble, PhysRevLett.120.077202, PhysRevLett.124.207201,PhysRevB.103.104408}. 
Importantly, experiments have successfully observed SkX for centrosymmetric triangular lattice magnets such as Gd$_2$PdSi$_3$,
accompanied by a pronounced topological Hall effect \cite{Kurumaji_Science2019,Khanh_Nakajima_Yu2020,Yoshimochi_Takagi_Ju2024}.
\begin{figure}[!t]
    \centering
    \includegraphics[width=0.91\linewidth]{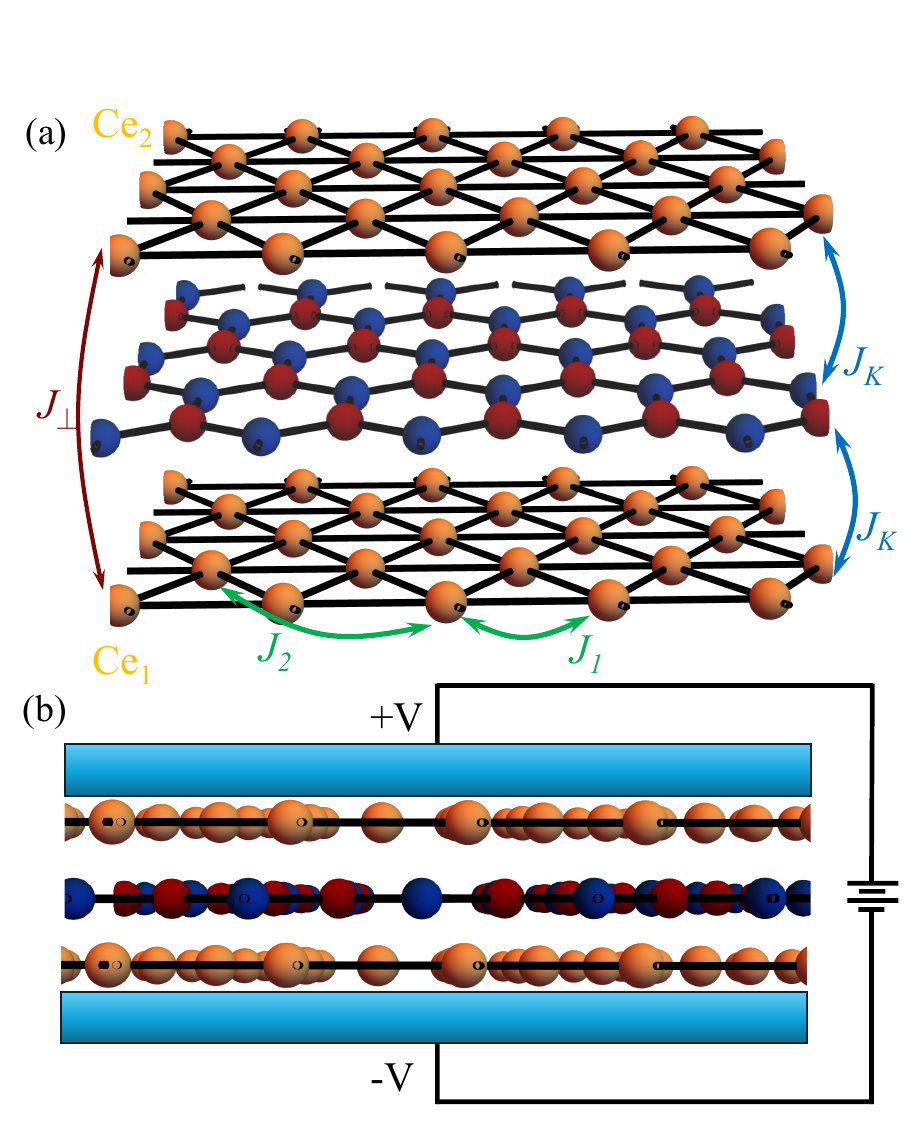}
    \caption{(a) Schematic of the model. In a monolayer of CeSiI, the cerium sites forms two layers of triangular lattices, indicated by Ce$_{1(2)}$. A silicene layer where the conduction electrons reside is sandwiched by the Ce layers. The local moments at Ce sites are coupled by intralayer NN ($J_1$) and NNN ($J_2$) Heisenberg interactions as well as an interlayer ($J_\perp$) interaction indicated by the arrows. J$_K$ indicates the Kondo coupling among the local moments and the conduction electrons. (b) Schematic of electric field tuning of a monolayer. Application of a potential difference increases (decreases) the Kondo coupling in layer 1(2).}
    \label{Fig:1}
\end{figure}
Recent neutron scattering experiments on CeSiI \cite{Okuma_PRR2021} show spiral magnetic order at $T_c =7.5$ K with an incommensurate ordering wave vector $\textbf{q}=(0.28,0)$ (r.l.u.) in the plane of the triangular lattice. This results in a short wavelength spiral of wavelength $L = |\textbf{b}|/|\textbf{q}|\approx 3.5$ (l.u.) where $\textbf{b}$ is a reciprocal lattice vector. The magnetic susceptibility shows highly anisotropic behaviour due to trigonal crystalline electric field splitting of the $J = 5/2$ manifold of the Ce$^{3+}$ $4f^1$ electrons. 
Consequently, a monolayer of CeSiI provides an ideal platform to study rich magnetic phases in triangular lattices with inversion symmetry with magnetic anistropy. Given the high tunability of the vdW materials, a monolayer of CeSiI or similar materials have the potential to stabilize topologically nontrivial spin textures, such as skyrmions.

Motivated by these advancements and the unique crystal structure of CeSiI, we construct an effective spin model with frustrated magnetic interactions and study its phase diagram using Landau-Lifshitz-Gilbert simulations implemented via the Sunny.jl package \cite{Sunny}. We tune the exchange parameters to reproduce the experimentally observed ordering wave vector in the absence of external magnetic field. Furthermore, we propose that the magnetic exchange interactions on each layer can be tuned by the application of an external electric field. Our main results are the following: (i) We find a large region in the phase diagram where SkX phase can be stabilized. (ii) Electric field allows for tuning of the phases on the two Ce layers, stabilizing the coexistence of different types of magnetic order on each layer. (iii) Electric field can lead to the stabilization of a vortex crystal phase (VtX) which can not be obtained for a monolayer otherwise.

The rest of the article is organized as follows. In Section II, we present the model and discuss how the exchange interactions are modified under applied electric field. In Section III, we present our results and discuss their implications. We conclude with a summary and an outlook in Section IV.

\begin{figure}[t]
    \centering
    \includegraphics[width=1\linewidth]{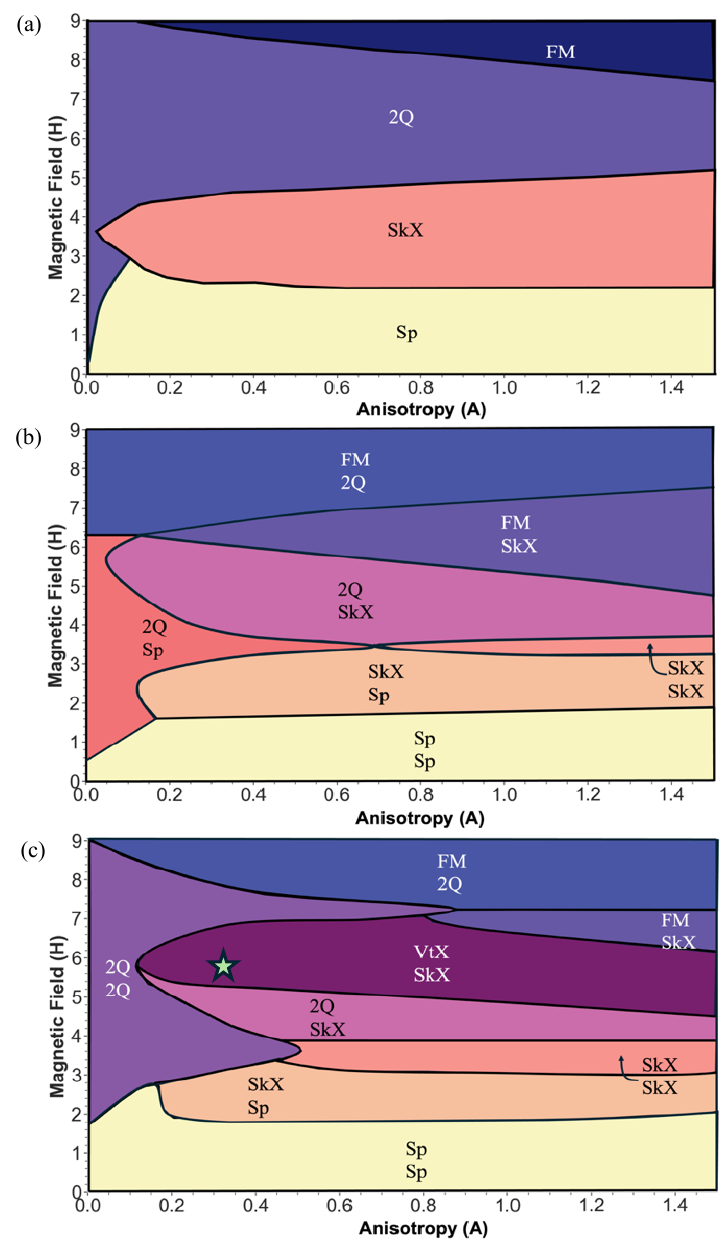}
    \caption{Ground state phase diagrams as a function of single site anisotropy (A) and external magnetic field (H). (a) In the absence of an external electric field ($\Delta V=0$). (b) For finite electric field ($\Delta V=0.2 \epsilon_f$/e), phase boundaries of the two layers shift, resulting in a phase diagram with coexisting phases ($J_\perp=0$).(c) Finite interlayer exchange $(J_{\perp 0}=-0.1)$ stabilizes symmetric phases and give rise to a new VtX-SkX phase. In Fig.~\ref{fig:3}, we present the real space configuration of the VtX-SkX state labelled by the star symbol.}
    \label{fig:2}
\end{figure}

\section{Model}
We consider an effective spin model defined on a bilayer triangular lattice (Fig.\ref{Fig:1}(a)),
\begin{eqnarray}
    H&=&\sum_{\nu, ij}J_{\nu, ij}\mathbf{S}_{\nu,i}\cdot \mathbf{S}_{\nu,j} + J_\perp \sum_i \mathbf{S}_{1,i}\cdot \mathbf{S}_{2,i}  \nonumber \\ && - A\sum_{\nu,i} (S_{\nu,i}^z)^2 -H\sum_{\nu,i}(S^z_{\nu,i})
    \label{eq:H}
\end{eqnarray}
where $\nu=1,2$ is the Ce layer index. The first term includes a ferromagnetic (FM) NN interaction ($J_1$) and an antiferromagnetic (AFM) NNN interaction ($J_2$). The second term is a FM interlayer interaction ($J_\perp)$. $A$ and $H$ are single site anisotropy and external magnetic field respectively. In order to determine the ratio of $J_2/J_1$, we perform a Luttinger-Tisza analysis \cite{Luttinger_PR1946} and minimize the exchange interaction in momentum space, $J(\textbf{q})=\sum_{ij}J_{ij}e^{i\textbf{q}\cdot (\textbf{r}_i-\textbf{r}_j)}$. The minima for the wave vector $\textbf{q}=(0.28,0)$ leads to a ratio of the exchange interactions $J_2/J_1= -1.6 $. The magnitude of the $J_2$ coupling being stronger than $J_1$ coupling may seem unrealistic, however we propose the following arguments to justify our choice: (i) In the long wavelength limit, the magnetic order is dictated by $J(q)$ around the order wave vector $\mathbf{q}$ that minimizes $J(q)$ \cite{lin2016ginzburg}. The detailed form $J(q)$ over the whole Brillouin zone does not play a role. The choice of Hamiltonian in Eq. \eqref{eq:H} is intended to reproduce the most relevant 
$J(q)$ around the ordering wave vector. (ii) if we consider an extended range model with an additional third nearest neighbor interaction \cite{Ivanov_JMMM1995}, the same wave vector may be realized with $J_2/J_1 = -1.05$ and $J_3/J_1 = 0.458$ (see Appendix A for details). We believe this fine tuning of parameters would not lead to any meaningful change in the analysis below. iii) The NN ferromagnetic Heisenberg coupling may be weaker as a result of a competition between the FM RKKY interaction mediated by the conduction electrons and the AFM superexchange between the nearest neighbours mediated by virtual hopping through unoccupied bands. While this competition may weaken $J_1$, it does not affect $J_2$, which in turn may lead to $J_2$ being stronger than $J_1$. First principles calculations suggest that the interlayer exchange is ferromagnetic ($J_\perp <0)$\cite{Posey_Nature2024}. For the remainder, we use units of $J_1 =-1$.

To break the layer symmetry and realize the coexistence of different magnetic phases on two layers, we explore the effects of external electric field. CeSiI is a heavy fermion material and it is anticipated that the magnetic exchange interactions originate from RKKY interactions. First principles calculations show that the conduction electrons predominantly reside at the silicine layer~\cite{Jang_npj2DMat2022, Fumega_NanoLett2024}. Therefore, the RKKY interaction arises via second order processes where the Ce local moments couple to the conduction electrons via the Kondo interaction~\cite{Vijayvargia_PRB2024}. We consider a setup where a potential $V$ and -$V$ is applied to the top and bottom gates as shown schematically in Fig.~\ref{Fig:1}(b). This doesn't affect the chemical potential of the conduction band. However, it changes the energy levels of the two cerium layers $\epsilon_{f1(2)}$, increase (decrease) with respect to the chemical potential due to the potential drop, $\epsilon_{f1/2} \rightarrow \epsilon_{f1/2}\mp e\Delta V$. Since Ce ions primarily fluctuate via $f^1 \leftrightarrow f^0$, the Kondo exchange on two layers gets modified as follows, $J_K^{1/2} = t^2_{cf}/(\epsilon_f\mp e\Delta V) = J_{K0}/ (1\mp e\Delta V/\epsilon_f)$ where $J_{K0}=t^2_{cf}/\epsilon_f$ is the Kondo coupling in the absence of an external electric field. Since $J_{RKKY} \sim J_K^2/W$ ($W\equiv$ bandwidth), the Heisenberg interaction will be modified as $J^{1/2}_{H}= J_{H0}/(1\mp e\Delta V/\epsilon_f)^2$. This implies that both NN and NNN Heisenberg couplings of layer $1(2)$ will increase(decrease) by a factor of $1/(1\mp e\Delta V/\epsilon_f)^2$ with an applied potential $V$. In addition, the interlayer Heisenberg interaction between the Ce local moments of two layers is also mediated via an RKKY mechanism which involves the Kondo coupling from each layer. As a result the interlayer exchange scales as $J_\perp \rightarrow  J_{\perp0}/ ((1+e\Delta V/\epsilon_f)(1-e\Delta V/\epsilon_f))$. Since the electric field only rescales all the intralayer couplings globally, the order wavevector remains the same.

\section{Results and discussion}
The ground state phase diagrams are obtained by performing Landau-Lifshitz-Gilbert simulations for the classical spin Hamiltonian (eq.~\ref{eq:H}) using the Sunny.jl package~\cite{Sunny}. The magnetic unit cell is approximately of $L = 1/0.28 \approx 3.5 $ unit cells. Hence, we consider $14 \times 14$ and $25 \times 25$ system sizes in our simulations. Our analysis indicates that our results are independent of the system size.
First, we explore the phase diagram in the absence of external electric field in Fig.~\ref{fig:2}(a). In this case, the interlayer exchange does not affect the phase diagram. Our results are qualitatively similar to other calculations with frustrated $J_1$-$J_2$ models on triangular lattices~\cite{leonov_natcomms2015,lin2016ginzburg}. We obtain single-q spiral phases labelled Sp for small H, SkX for intermediate H and double-q (\hz{2Q}) phases for larger H and smaller A. Note that we have combined the different single-q phases and double-q phases, as observed in the phase diagram in \cite{leonov_natcomms2015} into a single respective class. 

\begin{figure}[t]
    \centering
    \includegraphics[width=1\linewidth]{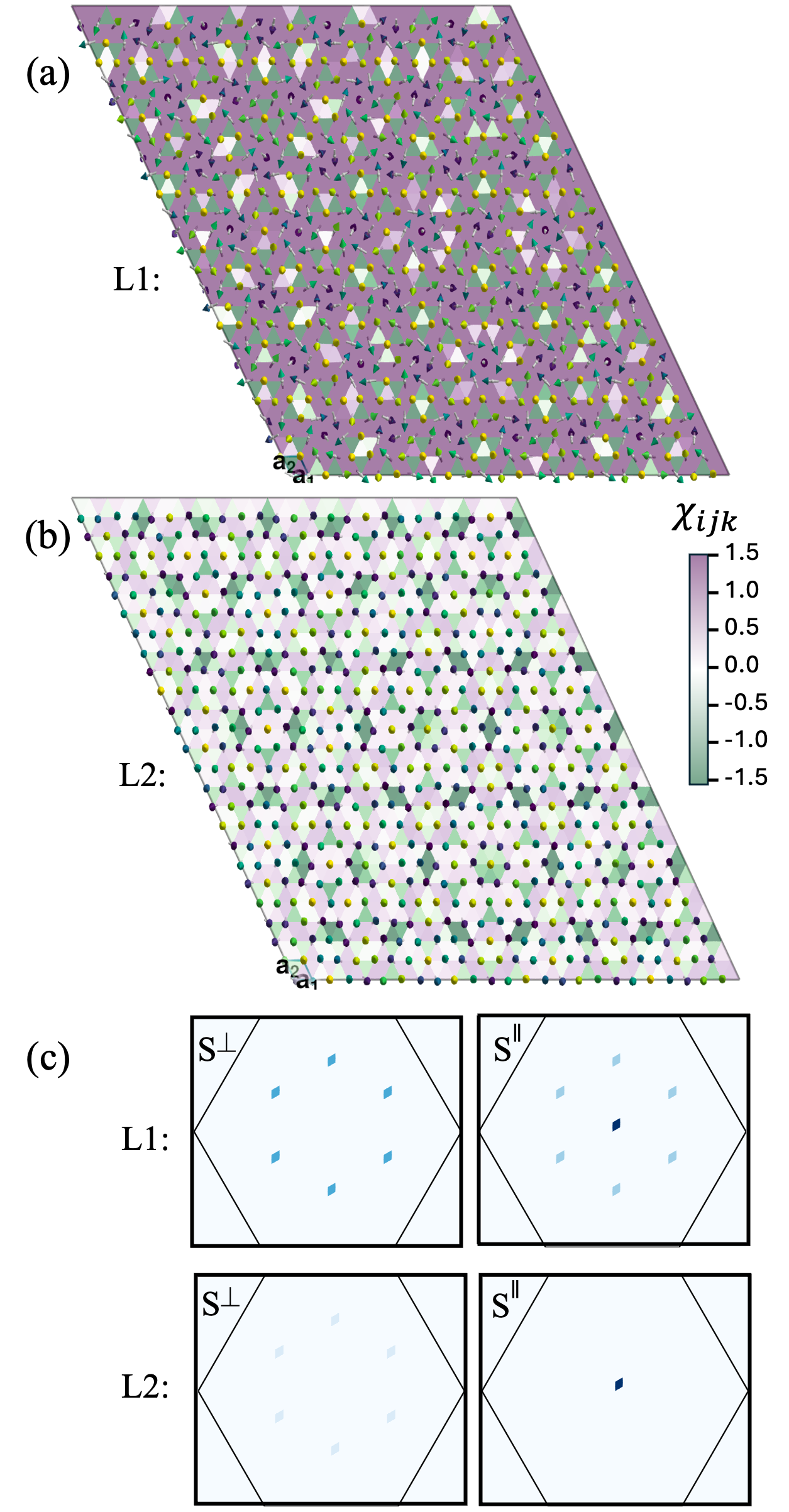}
    \caption{Real space spin textures corresponding to the phase VtX-SkX labelled by the stary symbol ($\star$) in Fig.~\ref{fig:2} (c). (a) Skyrmion crystal on layer 1 (L1) (b) Vortex crystal on layer 2 (L2) where $\chi_{ijk}$ is the local scalar chirality. (c) The intensity plot of the spin structure factor in the first BZ for the skyrmion lattice phase (L1) and the vortex lattice phase (L2). The left panels show the $xy$ components ($S^\perp$) and the right panels show the $z$ component ($S^\parallel$) of the spin structure constant.  }
    \label{fig:3}
\end{figure}

\begin{figure*}
    \centering
    \includegraphics[width=\textwidth]{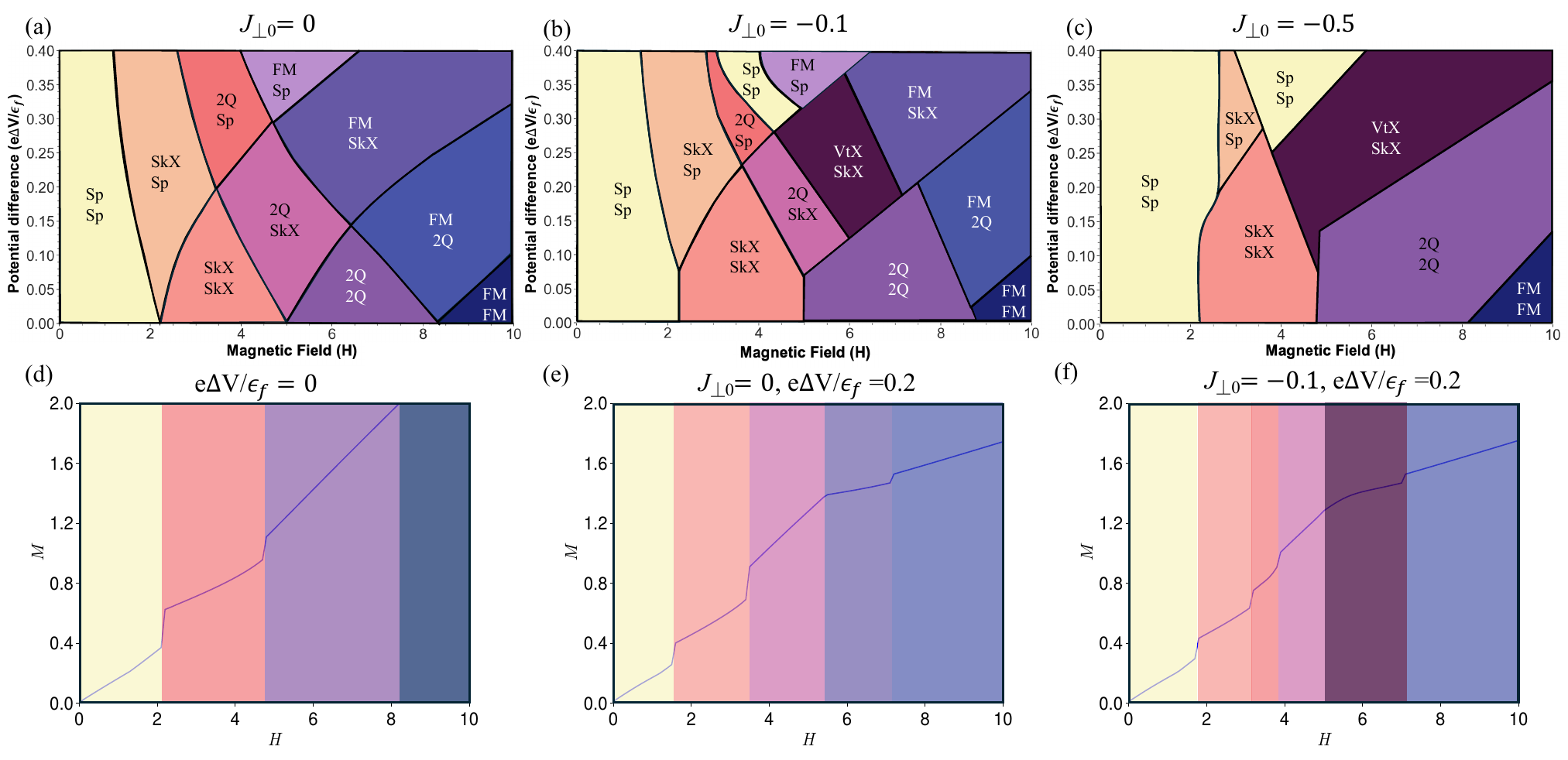}
    \caption{Phase diagrams as a function of electric and magnetic field for the interlayer exchange (a) $J_\perp = 0$, (b) $J_\perp = -0.1$, (c) $J_\perp = -0.5$. The single ion anisotropy is set to $A=0.8$. (a) Electric field breaks the layer symmetry and shift the phase boundaries for the two layers, leading to coexistence of different phases on the two layers. For non-zero interlayer interaction, a threshold electric field is required to stabilize coexistence of distinct phases on the two layers. A vortex crystal phase coexisting with skyrmion crystal is stabilized by the interlayer interaction. For higher interlayer interaction, the VtX-SkX phase expands. Magnetization as a function of magnetic field for (d) zero electric field ($\Delta V=0$) and  $\Delta V =0.2 \epsilon_f/e$ without (e) and with (f) interlayer interaction, exhibiting the coexistence phases and VtX-SkX phase. The phases are labelled by the corresponding colors shown in the phase diagram.}
    \label{fig:4}
\end{figure*}

Next we consider the effects of electric field in the absence of interlayer exchange ($J_{\perp 0}=0$).
Applying a potential difference as depicted in Fig.~\ref{Fig:1}(b) increases the Heisenberg exchange parameters of layer 1 (bottom) while decreasing them of layer 2 (top), keeping the on-site anisotropy constant. This leads to shifting of phase boundaries in the anisotropy-magnetic field phase diagram Fig.~\ref{fig:2}(a) for each layer differently. In Fig.~\ref{fig:2}(b) we present the phase diagram for the bilayer with the potential difference $\Delta V= 0.2 \epsilon_f/e$. Except the small regions where the two layers have the same phase, the majority of the phase diagram is occupied with regions where different phases on the two layers coexist. This is expected, since the first layer with the strengthened Heisenberg exchanges has a more stable spiral phase, hence covering more of the phase diagram, on the other hand the second layer has weaker Heisenberg exchanges relative to the anisotropy and the applied magnetic fields, resulting in a much bigger FM region in the phase diagram. Note that the regions where both the layers have same phases have shrunk considerably. 

Introducing a ferromagnetic interlayer interaction, with $J_{\perp 0}=-0.1$, we note a few changes in the phase diagram as presented in Fig.~\ref{fig:2}(c). Firstly, the stability of symmetric phase on two layers increases, as the FM interlayer exchange lowers the energy when the spin configuration on the two layers are identical. Secondly, we observe a vortex crystal phase on the second layer coexisting with a skyrmion phase on the first layer. VtX phase is also a 3Q spin texture, similar to a skyrmion. However, the difference between the two lies in the important observation that the topological charge of a VtX is zero ($\pi_2(S^2)=0$). The spin vector wraps some fraction of the sphere around the north pole, up to some threshold radius but never the whole sphere. However, because of the chirality of the $xy$-spin components, they can be categorized as Abelian vortices $\pi_1(S^1)$~\cite{kamiya2014vortex}. The real space spin texture of the SkX-VtX phase is presented in Fig.~\ref{fig:3} along with the corresponding spin structure factor.

In order to investigate the effects of electric field and the emergence and stability of the vortex crystal phase further, we fix the anisotropy to $A=0.8$ and present phase diagrams in the as a function of external electric and magnetic field for $J_{\perp 0}=0,-0.1$ and $-0.5$ in Fig.~\ref{fig:4}. Fig.~\ref{fig:4}(a) shows that the coexistence of different phases immediately appear for finite electric field. This effect is similar to what we observed as a function of $H$ and $A$ in Fig.~\ref{fig:2}(b). However, the interlayer interaction is minimized when the two layers have the same spin texture. For instance, Fig.~\ref{fig:4}(b) shows that a threshold electric field is required for the Sp-Sp and SkX-SkX phases to split into SkX-Sp while with no interlayer interaction Fig.~\ref{fig:4}(a), the splitting occurs with infinitesimal electric field. When examining the two phase diagrams in Fig.~\ref{fig:4}(a) and Fig.~\ref{fig:4}(b), while the phase boundaries are vaguely similar, regions with same phase on the two layers are more stable in Fig.~\ref{fig:4}(b) due to the interlayer interaction. The new phase VtX-SkX emerges in place of FM-SkX and some part of 2Q-SkX. It is observed that a 3Q SkX phase is necessary to stabilise the VtX phase in the other layer and this phase is connected to the FM-SkX phase. 

In Fig.~\ref{fig:4}(d), we present the magnetization as a function of $H$ for $\Delta V=0$ which shows a single-q phase, a skyrmion phase followed by a 2-q phase and a fully polarized FM phase. Similar behavior with a magnetization plateau at intermediate field has been observed in CeSiI \cite{Posey_Nature2024} as well as in Gd$_2$PdSi$_3$ \cite{Kurumaji_Science2019} which also possesses skyrmions. Furthermore, magnetization curves for $\Delta V = 0.2 \epsilon_f/e$ are plotted for $J_\perp = 0$ and $-0.1$ in Fig.~\ref{fig:4}(e,f). It can be seen that the jumps between the phases are smaller and more in number compared to the zero electric field case as the layer symmetry is broken in presence of electric field and the phase transitions in the two fields happen at different external fields. 

For higher values of interlayer exchange ($J_\perp = -0.5$), the stability of different phases coexisting decreases as shown in Fig.~\ref{fig:4}(c). This is because the energy cost of having unaligned spin configurations increases with the interlayer interaction. However, we observe that the VtX-SkX phase gets more stable with increasing interlayer interaction. Another feature to note is the resurgence of the Sp-Sp phase for high electric fields for intermediate magnetic fields. This happens because interlayer interaction also strengthens as the electric field is increased, making the breaking of layer symmetry cost more energy. The appearance of the SkX-VtX can be understood by considering a large potential difference regime as shown in Fig. \ref{fig:4} (c). In this regime, the spins in the layer with a much stronger $J_{ij}$ serve as an effective magnetic field Zeeman coupled to the spins in the layer with weaker  $J_{ij}$s. The skyrmion spin texture produces a nonuniform magnetic field, which can stabilize the VtX.


We established how electric field tuning can be used to tune the Kondo coupling in vdW heavy Fermi liquid platforms. In CeSiI, the conduction electrons reside on the silicene layer, consequently the chemical potential isn't affected by the applied electrical field. Importantly, this spatial separation of the conduction electrons and local moments leads to an energy difference between the two, resulting in the tuning of Kondo coupling. Such a spatial separation of conduction electrons and local moments in a vdW material can allow for an easy tuning knob for the Kondo coupling by applying an electric field along the direction of separation. While the effects of external electric field depends on the crystal structure of the materials, the proposed mechanism can be adapted for different materials. For instance a Kondo lattice has been proposed for vdW magnet CsCr$_6$Sb$_6$\cite{Song_arXiv2024}.

\section{Conclusion}
We explored the phase diagram of a bilayer triangular lattice spin model inspired by the vdW heavy fermion material CeSiI. We demonstrated that application of an electric field can tune the Kondo coupling differently for each Ce layer, allowing for adjustments of magnetic interactions. We showed that this tunability leads to the coexistence of different magnetic orders within a single monolayer. We investigated the phase diagrams as a function of anisotropy, magnetic field, and electric field. Our results show that electric field tuning can stabilize a new phase which is a vortex crystal coexisting with a skyrmion crystal. We also examined the role of interlayer interactions in stabilizing these chiral states. The ability to fine-tune magnetic interactions using electric fields provides a new avenue for controlling complex spin textures. Interesting future directions include electric field tuning other phases such as superconductivity in heavy fermion vdW materials and their moir\'e superlattices. Furthermore, a particularly interesting aspect of the disparity in magnetic interactions due to applied electric field is the current driven dynamics of the spin texture. Due to the difference in the Kondo coupling between the spins in the top and bottom layers to the conduction electron in the middle layer, the spin transfer torque for the spins induced by the electric current in the middle layer is also different between two layers. This can lead to a dynamical decoupling between the spin textures in the two layers for a large current. We leave this possibility for future exploration.

\section{Acknowledgements}
We thank Cristian Batista and  Michael Ziebel for fruitful discussions. This work was supported by the U.S. Department of Energy, Office of Science, Office of Basic Energy Sciences,
Material Sciences and Engineering Division under Award Number DE-SC002524. The work at Los Alamos was carried out under the auspices of the U.S. DOE NNSA under contract No. 89233218CNA000001 through the LDRD Program, and was supported by the Center for Nonlinear Studies at LANL, and was performed, in part, at the Center for Integrated Nanotechnologies, an Office of Science User Facility operated for the U.S. DOE Office of Science, under user proposals $\#2018BU0010$ and $\#2018BU0083$. We thank ASU Research Computing Center for high-performance computing resources.

\appendix
\section{Monolayer effective spin model}
cerium local moments in a monolayer triangular lattice magnetically order with a wave vector $|\textbf{q}|=0.28$. To find an effective spin Hamiltonian which has a spiral ground state given by the above wave-vector, we employ an extended range Heisenberg model:

\begin{equation}
    H_{eff} = \sum_{\langle ij \rangle} J_{ij} \textbf{S}_i \cdot \textbf{S}_j
\end{equation}
where the first, second and third nearest neighbor couplings are denoted by $J_1$, $J_2$ and $J_3$ respectively.
\begin{figure}[!htbp]
    \centering
    \includegraphics[width=\linewidth]{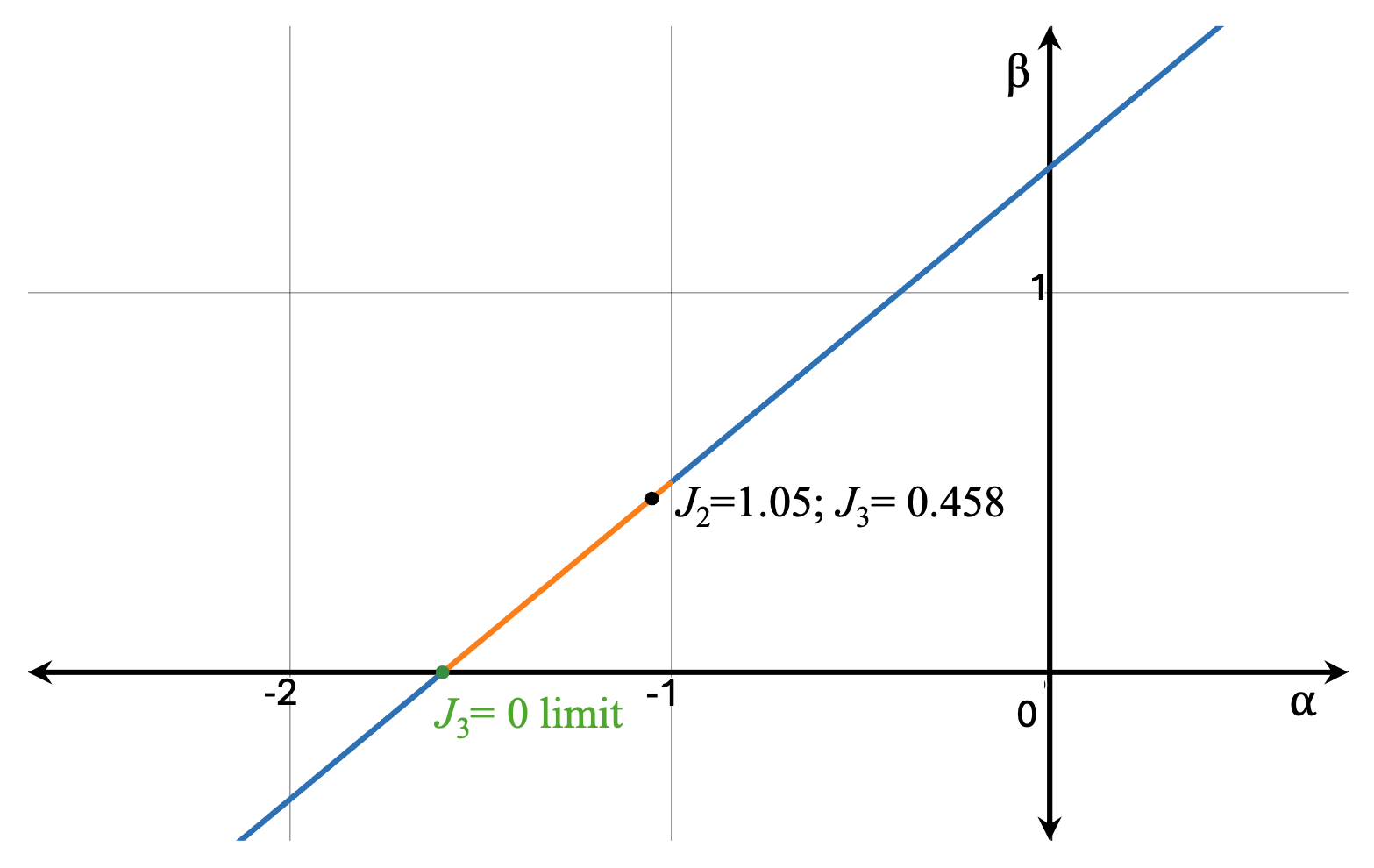}
    \caption{Allowed values of $\alpha=J_2/J_1$ and $\beta=J_3/J_1$, marked in yellow, that give rise to a spiral ground state with a wave vector $|\textbf{q}|=0.28 $ (r.l.u.). }
    \label{fig:A1}
\end{figure}
 After performing a linear spin wave approximation, one of the ground state solutions as a function of $\alpha \equiv J_2/J_1$ and $\beta \equiv J_3/J_1$ are found to be an incommensurate wave vector spiral. If the incommensurate wave-vector is along the reciprocal lattice vectors, say $\textbf{q} = (0,\ \pm Q_y)$, then we have \cite{Ivanov_JMMM1995}:
\begin{equation}
    \cos{\frac{\sqrt{3}Q_y}{2}} = -\frac{1+\alpha}{2\alpha +4\beta}
    \label{eq:J2J3}
\end{equation}

We fix $J_1=-1$ and find that the $J_3=0$ limit gives $\alpha = J_2/J_1=-1.6$. This is exactly what was obtained via Luttinger-Tisza analysis described in the main text. In Fig.~\ref{fig:A1}, we plot Eq.~\ref{eq:J2J3} for $Q_y=0.28 |b|$, where $|b|=4\pi/\sqrt{3}$ is the magnitude of the reciprocal lattice vectors. The range for which the solution gives an energy minima along a reciprocal lattice vector, is marked in yellow. 

Along the yellow segment, any values of $\{J_2, J_3 \}$ can be chosen and will lead to the same physics. We make a minimal choice by choosing the $J_3=0$ case, although that leads to a $J_2$ that is considerably larger than $J_1$. Another choice with a small $J_2$ value is $J_2=1.05$ and $J_3=-0.458$. However, as it does not change the outcome of our calculations, we choose to work in the $J_3=0$ limit.

\bibliography{references.bib}
\end{document}